\documentclass[preprint,12pt]{elsarticle}

\usepackage{amssymb}
\usepackage{color}
\usepackage{amsmath}
\usepackage{subfigure} 
\usepackage{CJK} 

\journal{X X X}

\begin{document}

\begin{CJK}{UTF8}{gbsn} 

\begin{frontmatter}



\title{ {\it Ab initio} Rayleigh-Schr\"{o}dinger perturbation calculation including three-body force }


\author{B. S. Hu (胡柏山)}
\author{T. Li (李通)}
\author{F. R. Xu (许甫荣)\corref{cor1}}

\address{State Key Laboratory of Nuclear Physics and Technology, School of Physics, Peking University, Beijing 100871, China}

\cortext[cor1]{frxu@pku.edu.cn}

\begin{abstract}
We first derive the Rayleigh-Schr\"{o}dinger many-body perturbation theory up to third order (RSPT3) for Hamiltonians with three-body interaction.
The structure of closed-shell nuclei in a wide mass range from $^4$He to $^{48}$Ca has been investigated by the RSPT3 with explicit $NN$+3$N$ Hamiltonian. The RSPT3 calculations are performed within Hartree-Fock bases.
The perturbative contribution of antisymmetrized Goldstone diagrams (diagrammatic expansion for RSPT) with normal-ordered interaction has been analyzed. 
We demonstrate that the normal-ordered two-body level (NO2B) approximation which neglects the residual three-body term can catch the main effect of full three-body force in RSPT calculation.
We also present rigorous benchmarks for RSPT3 with non-perturbative coupled cluster and in-medium similarity renormalization group  using the same chiral NNLO$_{\text{sat}}$ NO2B interaction. The three methods are in good agreement with each other for binding energies and radii.
However, the RSPT3 provides an alternative that is computationally inexpensive but comparable in accuracy to state-of-the-art non-perturbative many-body methods.
\end{abstract}


\end{frontmatter}

\end{CJK} 


\section{Introduction}
\label{sec:level1}
In recent years it has become clear that three-nucleon (3$N$) interaction plays an important role in nuclear matter and structure calculations from first principles \cite{Machleidt20111,PhysRevC.86.054317,PhysRevC.89.044321,PhysRevC.91.051301,PhysRevC.70.044005,
PhysRevLett.103.082501,PhysRevLett.106.202502,
PhysRevLett.110.022502,PhysRevLett.110.242501,PhysRevLett.113.262504,RevModPhys.87.1067}. 
When the degrees of freedom and the Hilbert space are restricted, the 3$N$ and higher-body (can be omitted) forces appear \cite{RevModPhys.85.197}.
On the one hand, describing the properties of atomic nuclei based on the fundamental interactions among protons and neutrons (i.e., nucleons) is a fundamental goal in nuclear theory. However, the nucleons are not point particle with virtual excitations and internal degrees of freedom \cite{PhysRevC.76.034302}. This leads to the initial 3$N$ force. On the other hand, a certain renormalization scheme is often used in practical calculation to deal with the strong short-range correlations of realistic force and speed up the convergence. This process will induce 3$N$ interaction.

For most {\it ab initio} many-body approaches, such as no-core shell model (NCSM), coupled-cluster (CC) or in-medium similarity renormalization group (IM-SRG) method,
the full inclusion of 3$N$ interaction is computationally very costly and often renders calculations impossible. In order to solve this dilemma, the normal-ordered two-body (NO2B) approximation to the full 3$N$ interaction has been widely used in nuclear structure calculations. In the NO2B approximation, the normal-ordered 3$N$ interaction in a chosen reference is truncated at the two-body level. Namely, the approximation includes the zero-, one-, and two-body parts of the 3$N$ interaction in normal-ordered form and neglects the residual normal-ordered three-body components.
The CC and importance-truncated no-core shell model (IT-NCSM) calculations \cite{PhysRevC.76.034302,PhysRevLett.109.052501} have shown that the normal-ordered three-body term can be neglected, and the NO2B approximation allows for accurate nuclear structure calculations for $NN$+3$N$ Hamiltonians.
It is worth probing that the contribution of different normal-ordered components of the three-body force in other many-body methods.

Rayleigh-Schr\"{o}dinger perturbation theory (RSPT) \cite{Rayleigh1894,Schrodinger1926,bartlett2009} is a powerful many-body theory which starts from a solvable mean-field problem and derives a correlated perturbed solution. However, it has been hindered for decades in nuclear calculations, because of the high non-perturbation of the realistic potential with strong short-range repulsion and strong tenser force \cite{PhysRevC.95.034321}.
Recently, with the development of modern renormalization technique and chiral effective field theory, the interaction can become more perturbative and ``soft". Therefore, many RSPT based methods are coming, such as RSPT with different renormalization schemes \cite{PhysRevC.68.034320,PhysRevC.69.034332,PhysRevC.73.044312,1674-1137-41-10-104101}, IT-NCSM \cite{PhysRevLett.99.092501} and Bogoliubov many-body perturbation theory \cite{Tichai:2018mll}. Therefore, more elaborate investigations of the RSPT are needed.
In our previous work \cite{PhysRevC.94.014303}, the two-body level RSPT had been developed in the Hartree-Fock (HF) basis within the angular momentum coupling representation.
We had successfully applied the developed RSPT to the structure of closed-shell nuclei.
It had been demonstrated \cite{TICHAI2016283,PhysRevC.94.014303} that in the HF basis the RSPT energy corrections up to third order (HF-RSPT3) can give well-converged results, while in the harmonic oscillator (HO) basis the corrections up to 30th order could be divergent even for softened interactions.
We also calculated the density correction and found that the second- and higher-order contribution to nuclear radius is not so remarkable. However, all the existing RSPT calculations lack three-body force up to third order.

In the present work, we investigate the application of explicit 3$N$ force in HF-RSPT3 and test the validity of the NO2B approximation to the full 3$N$ interaction. First, we formulate three-body level RSPT3 based on HF single-particle states. In order to reduce the computational task, the RSPT correction equations are transformed into the angular momentum coupled representation. Then, the new developed HF-RSPT3 are performed with three sets of $NN$+3$N$ Hamiltonian for the closed-shell structure calculation.
Finally, we use the state-of-the-art $\chi$NNLO$_{\text{sat}}$ interaction with NO2B approximation
to give a first-principle description of $^{4}$He, $^{14,22}$C, $^{16,22,24}$O and $^{40,48}$Ca.

\section{Theoretical framework}
\label{sec:1}
The intrinsic Hamiltonian of the $A$-nucleon system can be written as
\begin{equation}
\label{eq1}
\begin{split}
\hat{H} = &
\displaystyle\sum_{i=1}^{A} \left(1-\dfrac{1}{A}\right) \frac{\vec{p}_{i}^{2}}{2m} +
\displaystyle\sum_{i<j}^{A}   \left(\hat{V}_{NN,ij}-\frac{\vec{p}_{i}\cdot\vec{p}_{j} }{mA} \right) +
\displaystyle\sum_{i<j<k}^{A} \hat{V}_{3N,ijk}
\\
= & \displaystyle\sum_{i=1}^{A} {\hat{H}}^{(1)}_{i}+\displaystyle\sum_{i<j}^{A}{\hat{H}}^{(2)}_{ij}
+\displaystyle\sum_{i<j<k}^{A}{\hat{V}}_{3N,ijk},
\end{split}
\end{equation}
where $V_{NN,ij}$ is the two-body nucleon-nucleon ($NN$) interaction,
and $V_{NNN,ijk}$ is the 3$N$ interaction.
Three sets of $NN$+3$N$ Hamiltonians are adopted in this paper.
One is the $\chi$N$^{3}$LO \cite{PhysRevC.68.041001}
and JISP16 $NN$ interactions \cite{PhysRevC.70.044005,Shirokov200596,Shirokov200733}
plus spin-isospin-independent contact 3$N$ force \cite{PhysRevC.82.024319}.
Another one is the state-of-the-art $\chi$NNLO$_{\rm sat}$ \cite{PhysRevC.91.051301,PhysRevC.91.044001},
in which the $NN$ and 3$N$ interactions were
optimized simultaneously to low-energy nucleon-nucleon scattering data, as well as binding
energies and radii of selected nuclei up to $^{25}$O.
The phenomenological spin-isospin-independent contact 3$N$ interaction \cite{PhysRevC.82.024319} reads
\begin{equation}	
\label{threebody contact coordinate}
\hat{V}^{\text{ct}}_{3N}=C_{3N} \delta^{(3)}\left(\vec{x}_{1}-\vec{x}_{2}\right)
\delta^{(3)}\left(\vec{x}_{1}-\vec{x}_{3}\right)
\end{equation}
with variable strength $C_{3N}$.

In this work, we perform the closed-shell HF calculation with the established Hamiltonian (\ref{eq1}).
In spherical case, the angular momentum is preserved.
In what follows it is assumed that the indices $i$, $j$, $k$, ... label the occupied orbital (i.e., hole states) in the HF basis,
$a$, $b$, $c$, ... refer to the unoccupied orbital (i.e. particle states), and $p$, $q$, $r$, ... are any orbital (either hole or particle).
After the HF iteration, we can rewrite the Hamiltonian (\ref{eq1}) in a normal-ordered form
by choosing the HF Slater determinant $|\phi \rangle$ as the reference state,
\begin{equation}	
\label{Hwick}
\begin{split}
\displaystyle\hat{H}=&\sum_{i}\langle i|\hat{H}^{(1)}|i \rangle
\displaystyle+\frac{1}{2}\sum_{ij}\langle ij|\hat{H}^{(2)}|ij \rangle
\displaystyle+\frac{1}{6}\sum_{ijk}\langle ijk|\hat{V}_{3N}|ijk \rangle
\\
\displaystyle&+\sum_{pq}\left(\langle p|\hat{H}^{(1)}|q \rangle
\displaystyle+\sum_{i}\langle pi|\hat{H}^{(2)}|qi \rangle
\displaystyle+\frac{1}{2}\sum_{ij}\langle pij|\hat{V}_{3N}|qij \rangle \right):\hat{p}^{\dagger}\hat{q}:
\\
\displaystyle&+\frac{1}{4}\sum_{pqrs}\left(\langle pq|\hat{H}^{(2)}|rs \rangle
\displaystyle+\sum_{i}\langle pqi|\hat{V}_{3N}|rsi \rangle \right) :\hat{p}^{\dagger}\hat{q}^{\dagger}\hat{s}\hat{r}:
\\
\displaystyle&+\frac{1}{36}\sum_{pqrstu}\langle pqr|\hat{V}_{3N}|stu \rangle:\hat{p}^{\dagger}\hat{q}^{\dagger}\hat{r}^{\dagger}\hat{u}\hat{t}\hat{s}:
\end{split},
\end{equation}
where $:\hat{p}^{\dagger}\hat{q}:$ indicates the normally ordered product of the creation and annihilation operators.
As mentioned above, we take the HF Hamiltonian ($\hat{H}_{\rm HF}$) as a zero-order Hamiltonian in RSPT,
and the $A$-nucleon Hamiltonian (\ref{eq1}) can be separated into a zero-order part $\hat{H}_{\rm HF}$ and a perturbation $V$,
\begin{equation}
\hat{H} = \hat{H}_{\rm HF} + \hat{V}
\end{equation}
with
\begin{equation}	
\label{HFHamiltonianWick}
\begin{split}
\displaystyle\hat{H}_{\text{HF}}=&\sum_{i}\langle i|\hat{H}^{(1)}|i \rangle
\displaystyle+\frac{1}{2}\sum_{i,j}\langle ij|\hat{H}^{(2)}|ij \rangle
\displaystyle+\frac{1}{6}\sum_{i,j,k}\langle ijk|\hat{V}_{3N}|ijk \rangle
\\
\displaystyle&+\sum_{p,q}\left(\langle p|\hat{H}^{(1)}|q \rangle
\displaystyle+\sum_{i}\langle pi|\hat{H}^{(2)}|qi \rangle
\displaystyle+\frac{1}{2}\sum_{i,j}\langle pij|\hat{V}_{3N}|qij \rangle \right):\hat{p}^{\dagger}\hat{q}:
\end{split},
\end{equation}
and perturbation  part
\begin{equation}	
\label{NewperturbationV}
\hat{V}
=\displaystyle\frac{1}{4}\sum_{pqrs} \langle pq |\hat{W}| rs \rangle:\hat{p}^{\dagger}\hat{q}^{\dagger}\hat{s}\hat{r}:
\displaystyle+\frac{1}{36}\!\sum_{pqrstu}\langle pqr|\hat{V}_{3N}|stu
\rangle:\hat{p}^{\dagger}\hat{q}^{\dagger}\hat{r}^{\dagger}\hat{u}\hat{t}\hat{s}:,
\end{equation}
where
\begin{equation}	
\label{DefineW}
\langle pq |\hat{W}| rs \rangle=
\langle pq |\hat{H}^{(2)}| rs \rangle +
\sum_{i} \langle pqi | \hat{V}_{3N}| rsi \rangle.
\end{equation}
In NO2B approximation the last term in Eq.~(\ref{NewperturbationV}) is neglected,
and we take the $\hat{W}$ term as the perturbation part.

\begin{figure}
\centering
\setlength{\fboxrule}{0.3pt}
\setlength{\fboxsep}{0.10cm}
\fbox{
\shortstack[c]{
\subfigure[]
{
\includegraphics[width=1.00in]{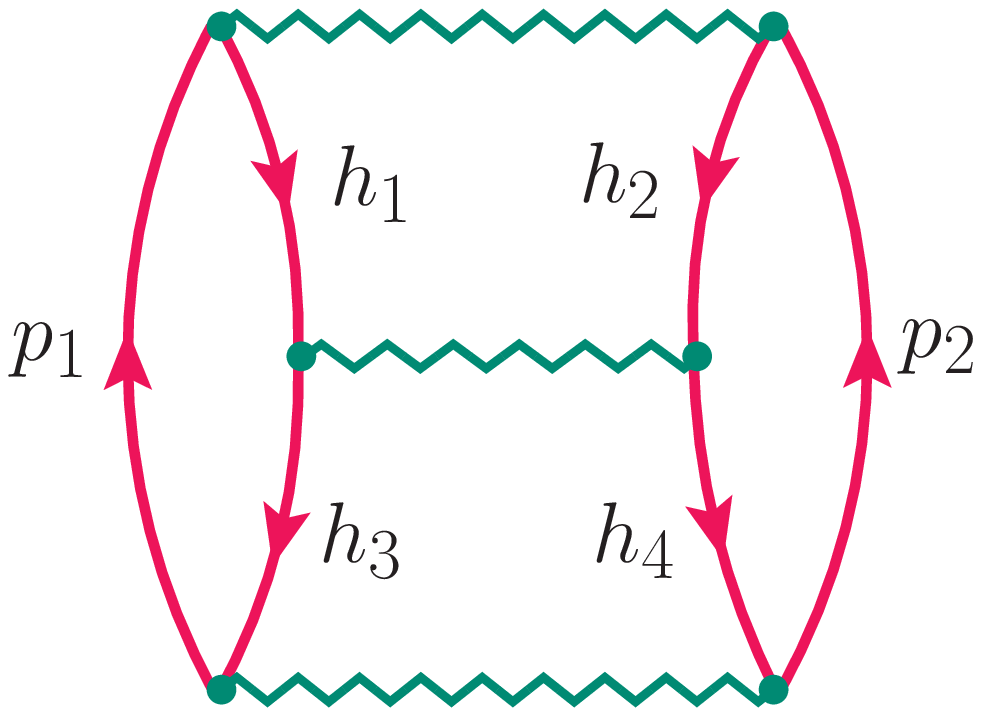}
}
\subfigure[]
{
\includegraphics[width=1.00in]{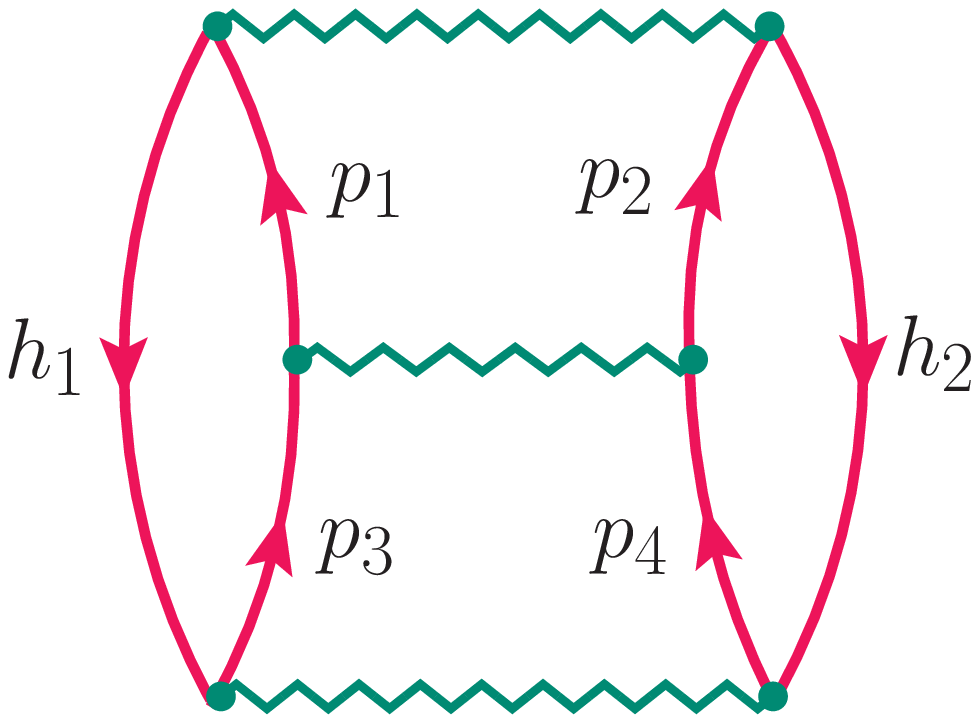}
}
\subfigure[]
{
\includegraphics[width=1.00in]{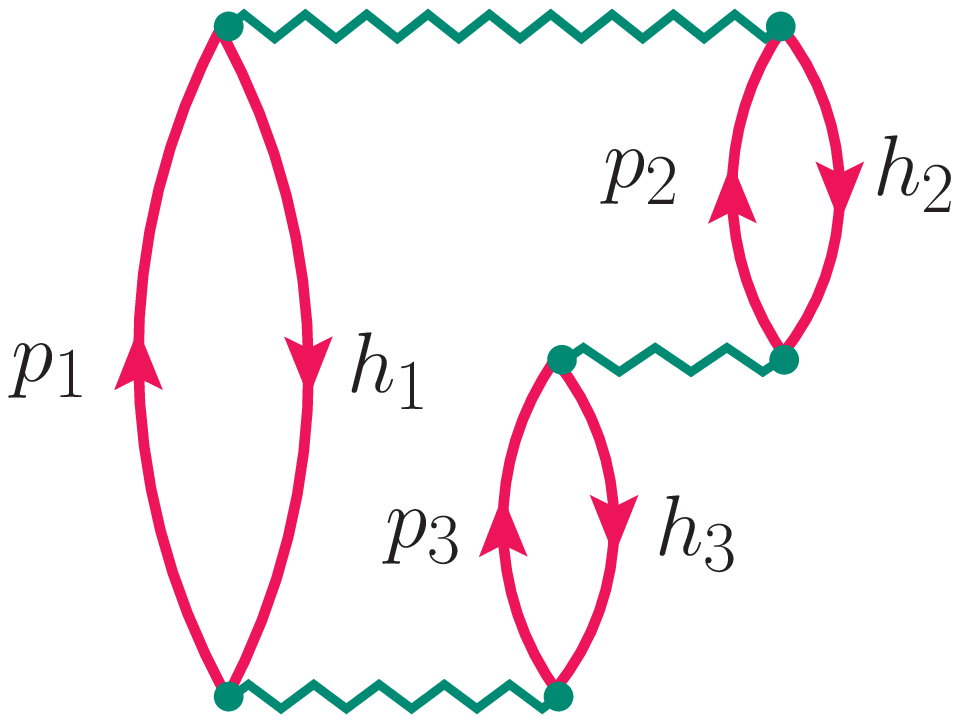}
}
\\
\subfigure[]
{
\includegraphics[width=1.20in]{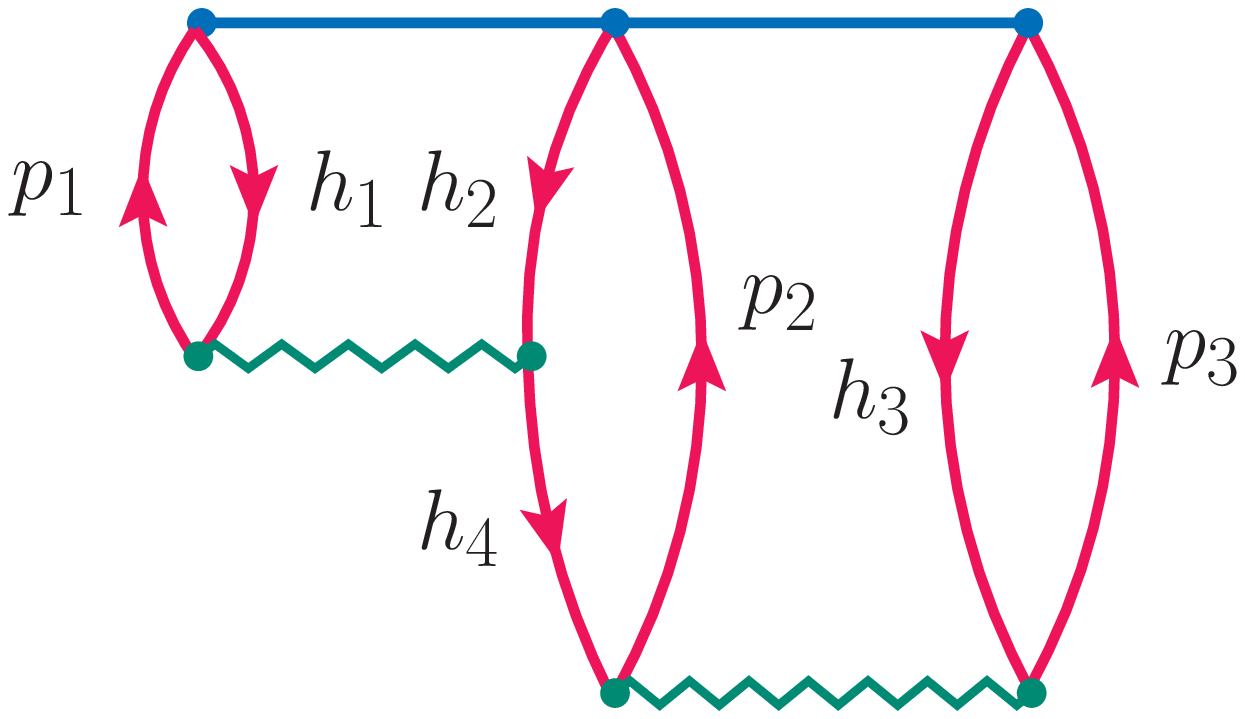}
}
\subfigure[]
{
\includegraphics[width=1.20in]{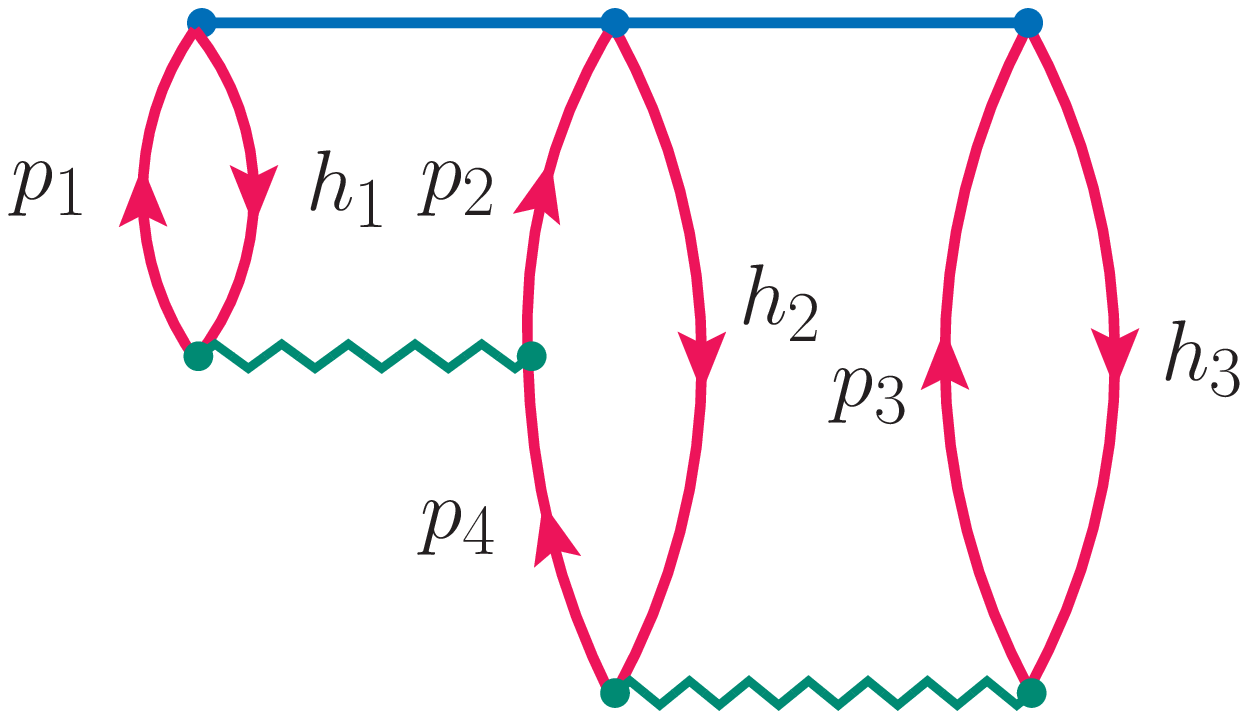}
}
\subfigure[]
{
\includegraphics[width=0.96in]{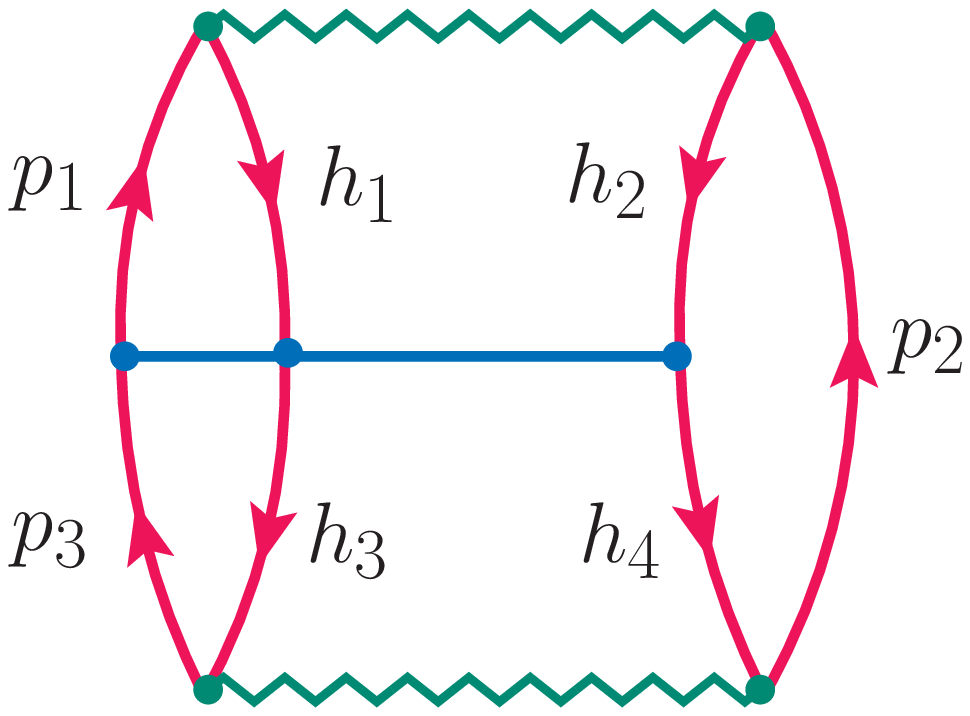}
}
\subfigure[]
{
\includegraphics[width=0.96in]{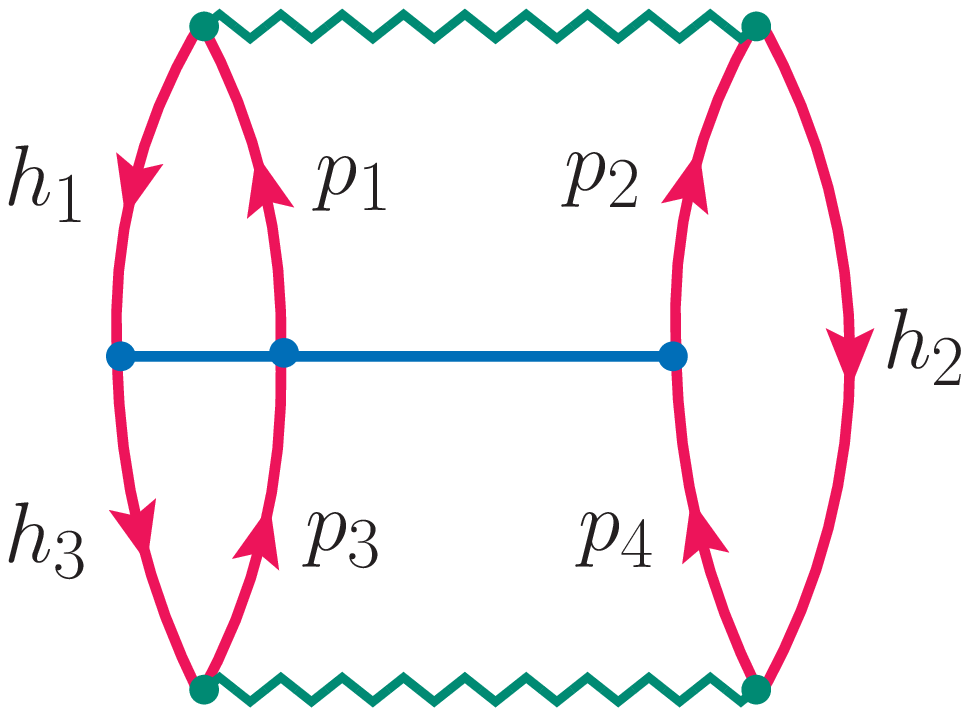}
}
\\
\subfigure[]
{
\includegraphics[width=1.16in]{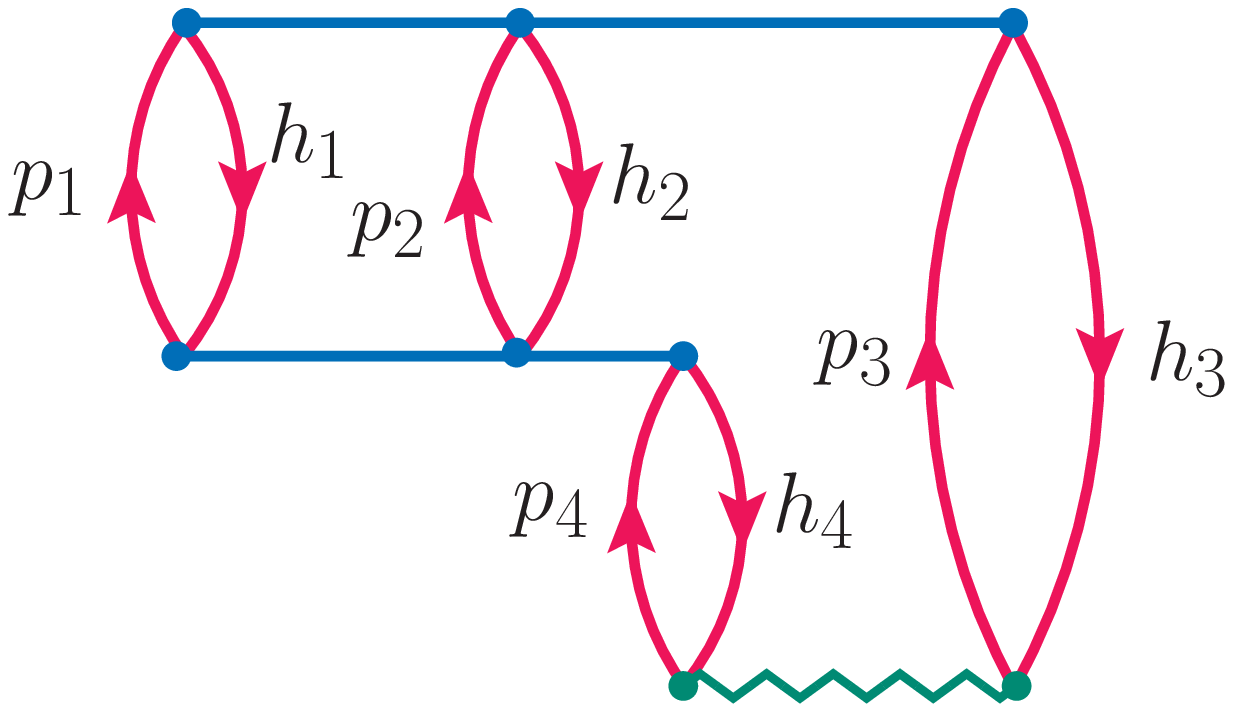}
}
\subfigure[]
{
\includegraphics[width=1.18in]{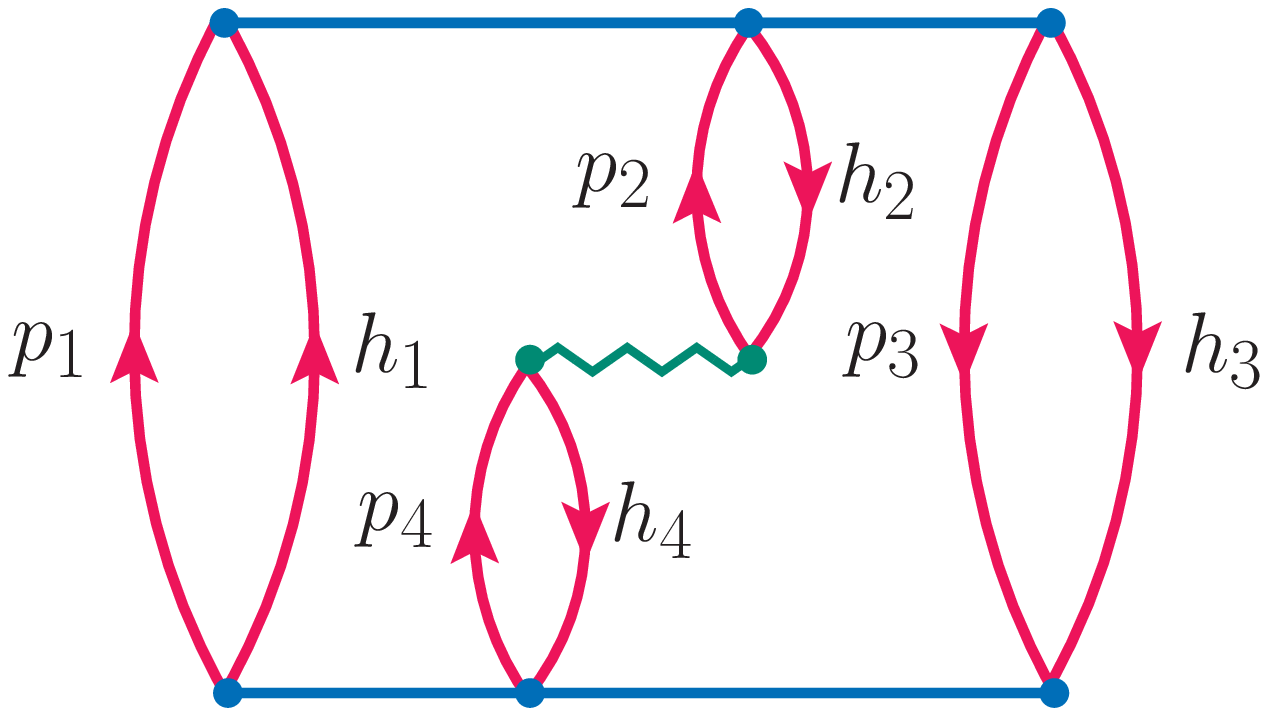}
}
\subfigure[]
{
\includegraphics[width=1.18in]{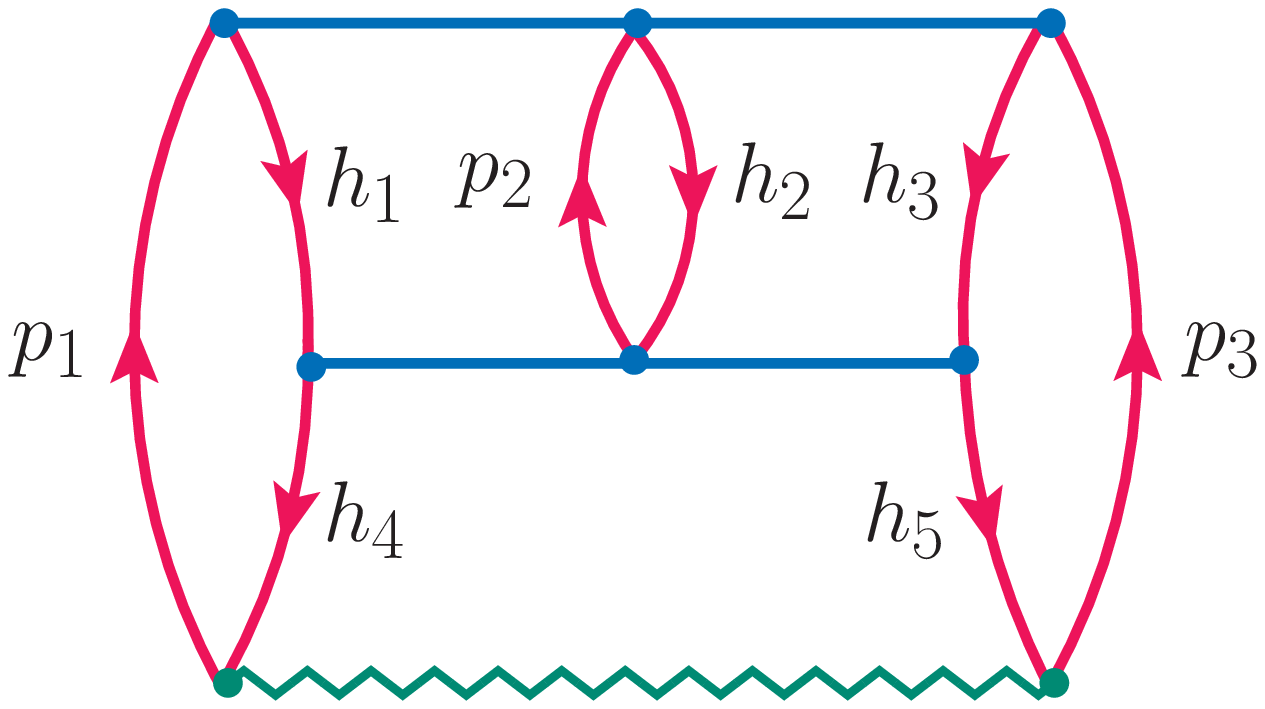}
}
\\
\subfigure[]
{
\includegraphics[width=1.18in]{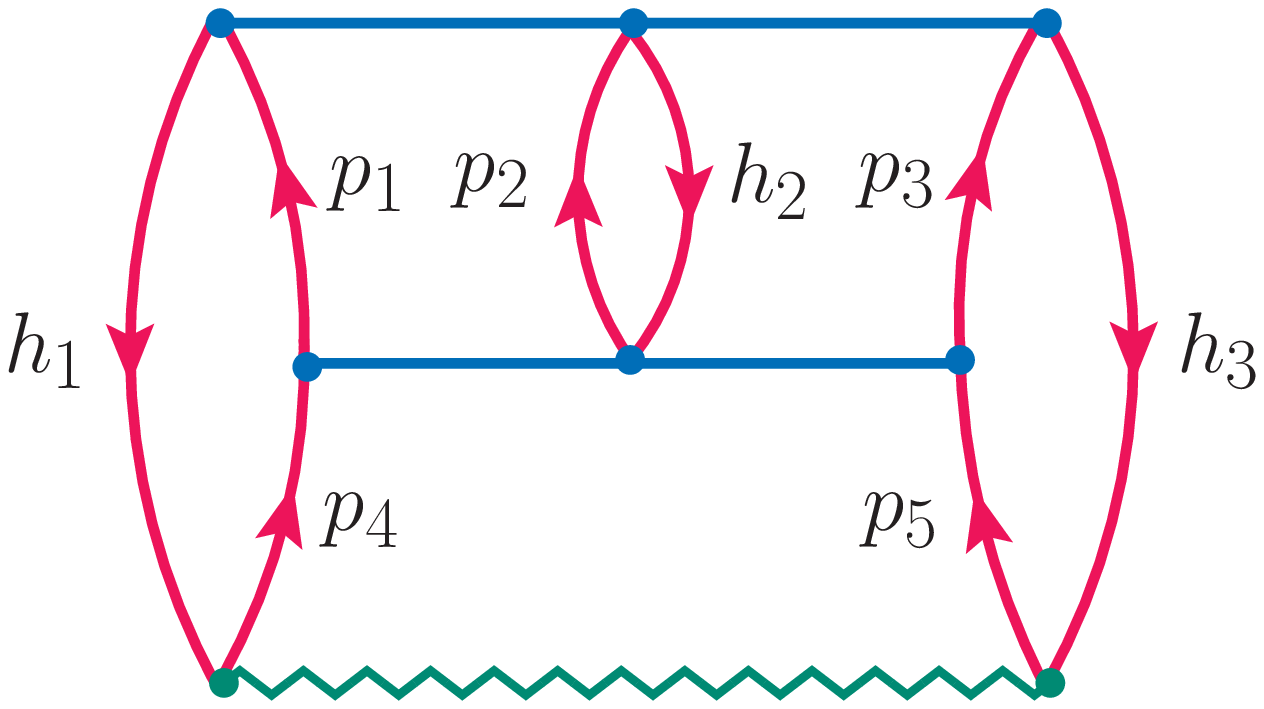}
}
\subfigure[]
{
\includegraphics[width=1.18in]{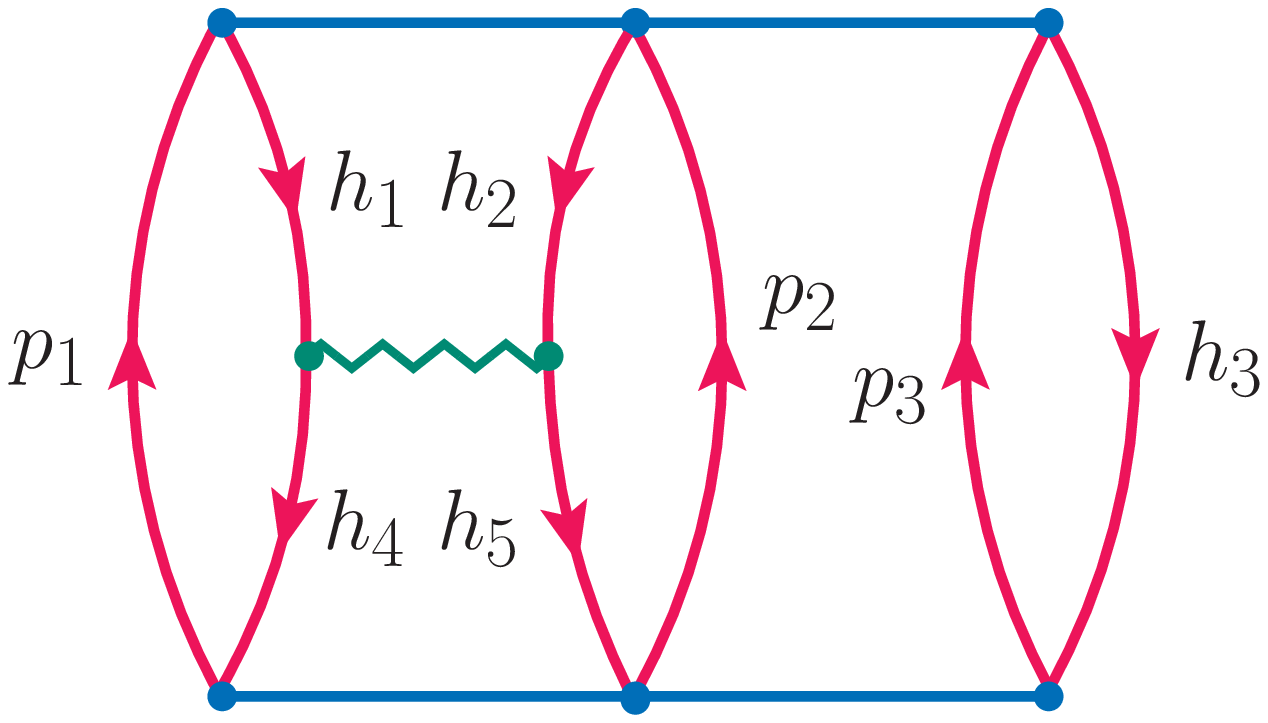}
}
\subfigure[]
{
\includegraphics[width=1.18in]{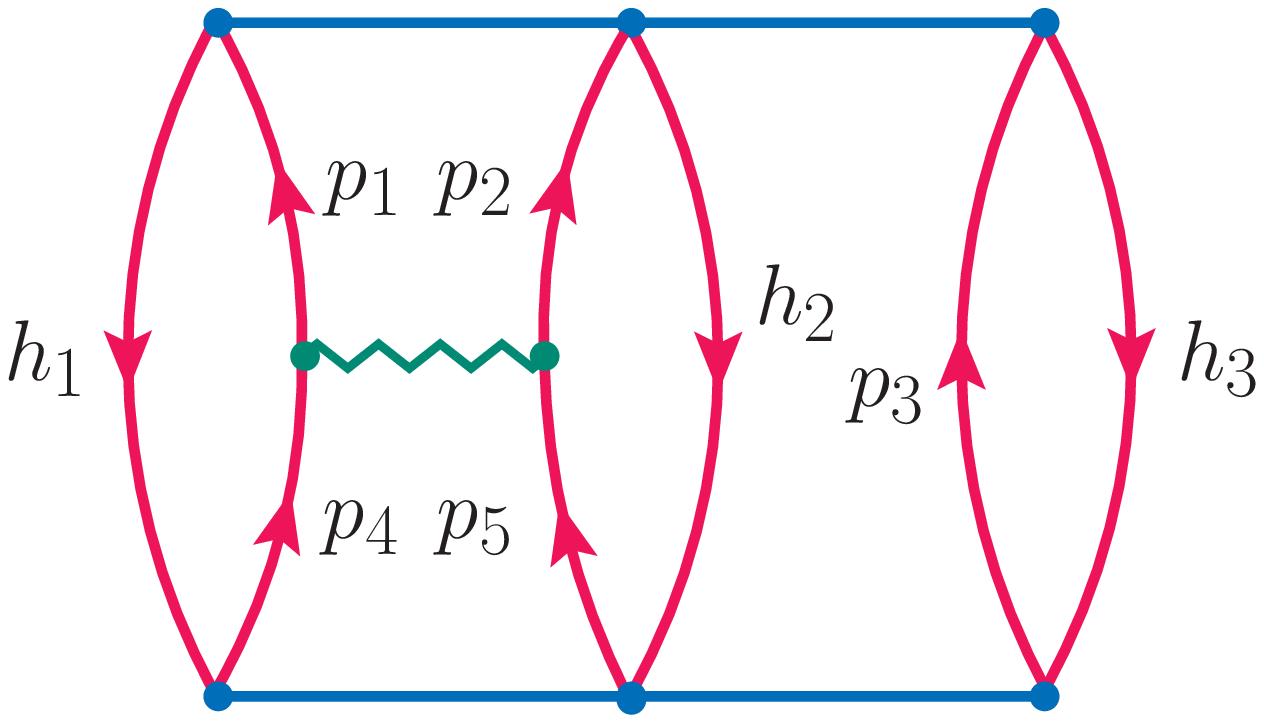}
}
\\
\subfigure[]
{
\includegraphics[width=1.15in]{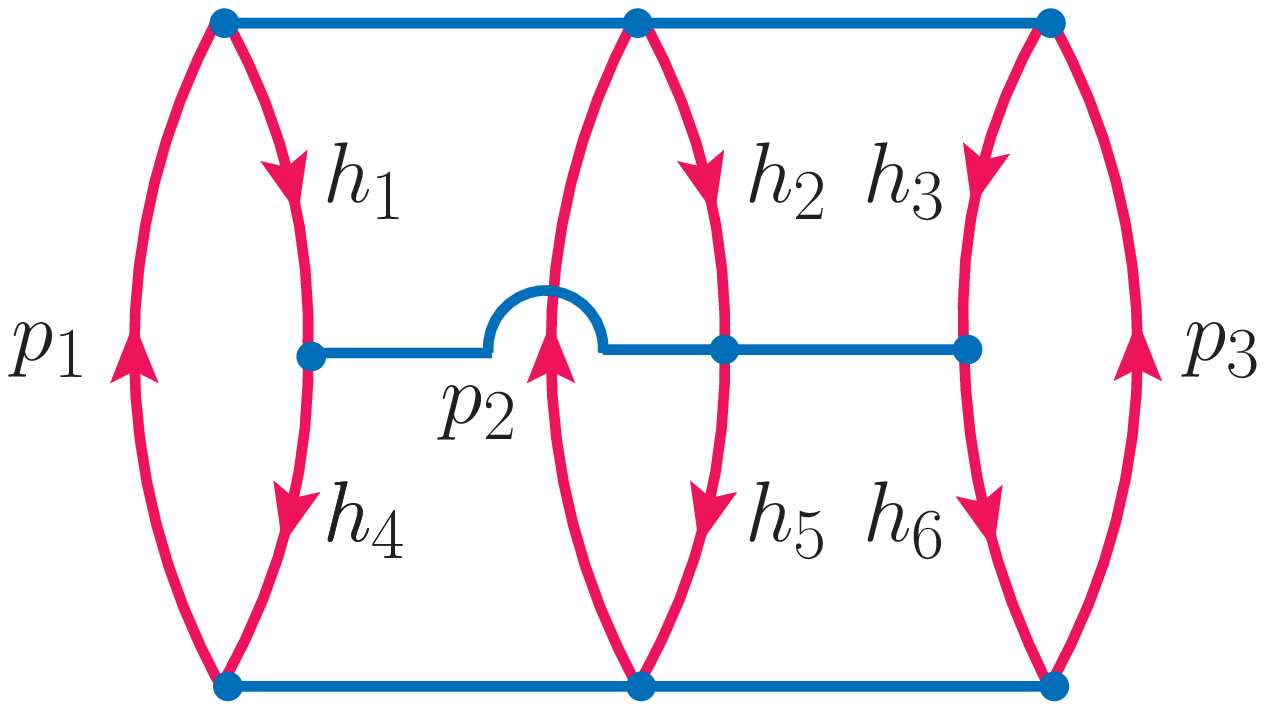}
}
\subfigure[]
{
\includegraphics[width=1.15in]{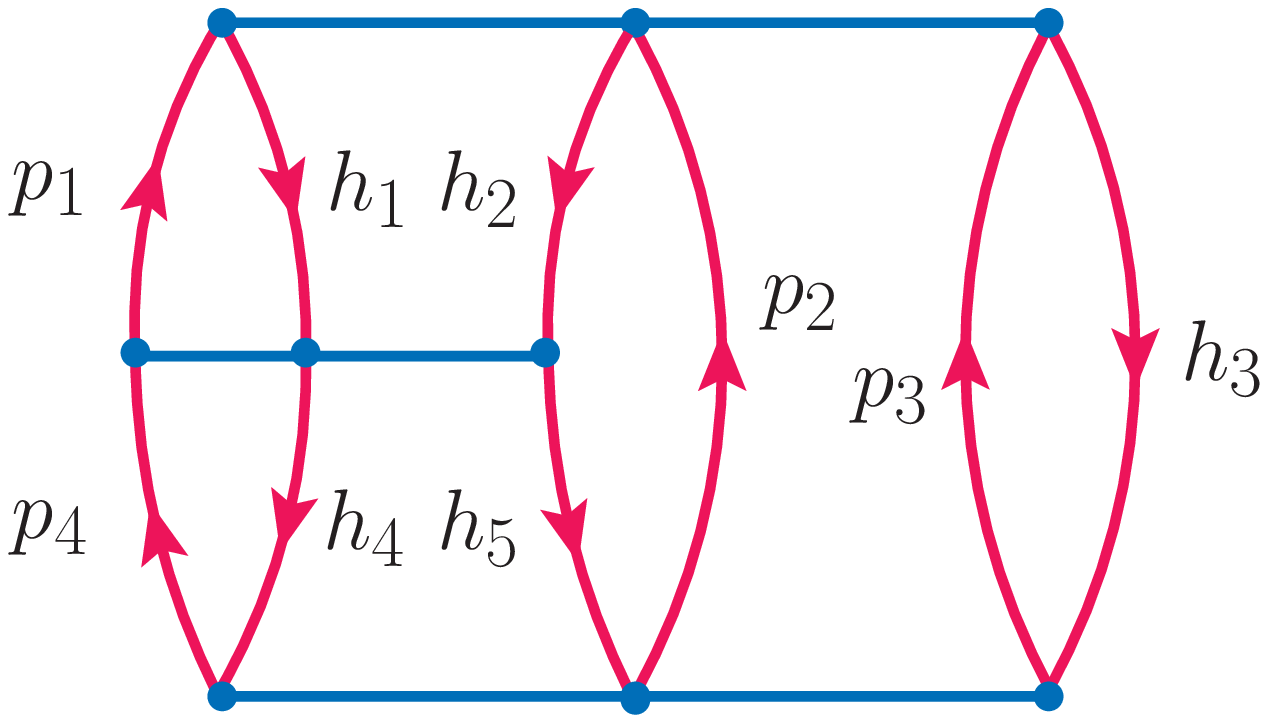}
}
\subfigure[]
{
\includegraphics[width=1.15in]{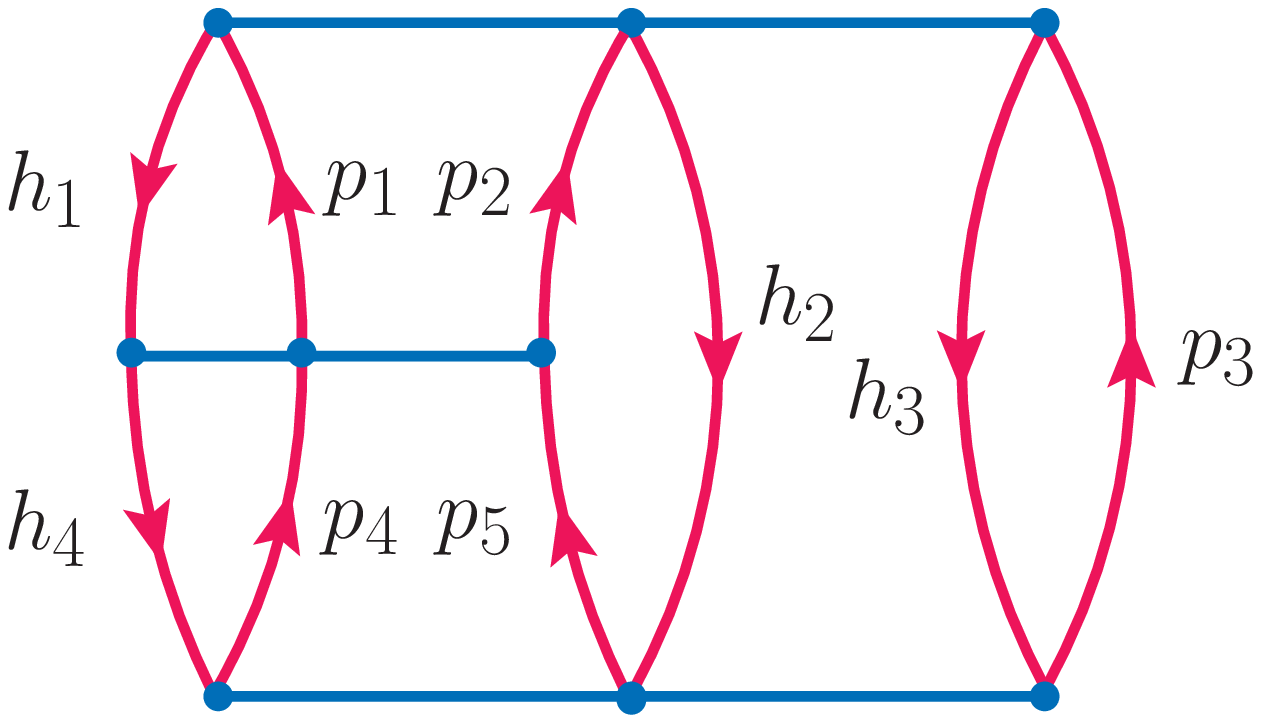}
}
\subfigure[]
{
\includegraphics[width=1.15in]{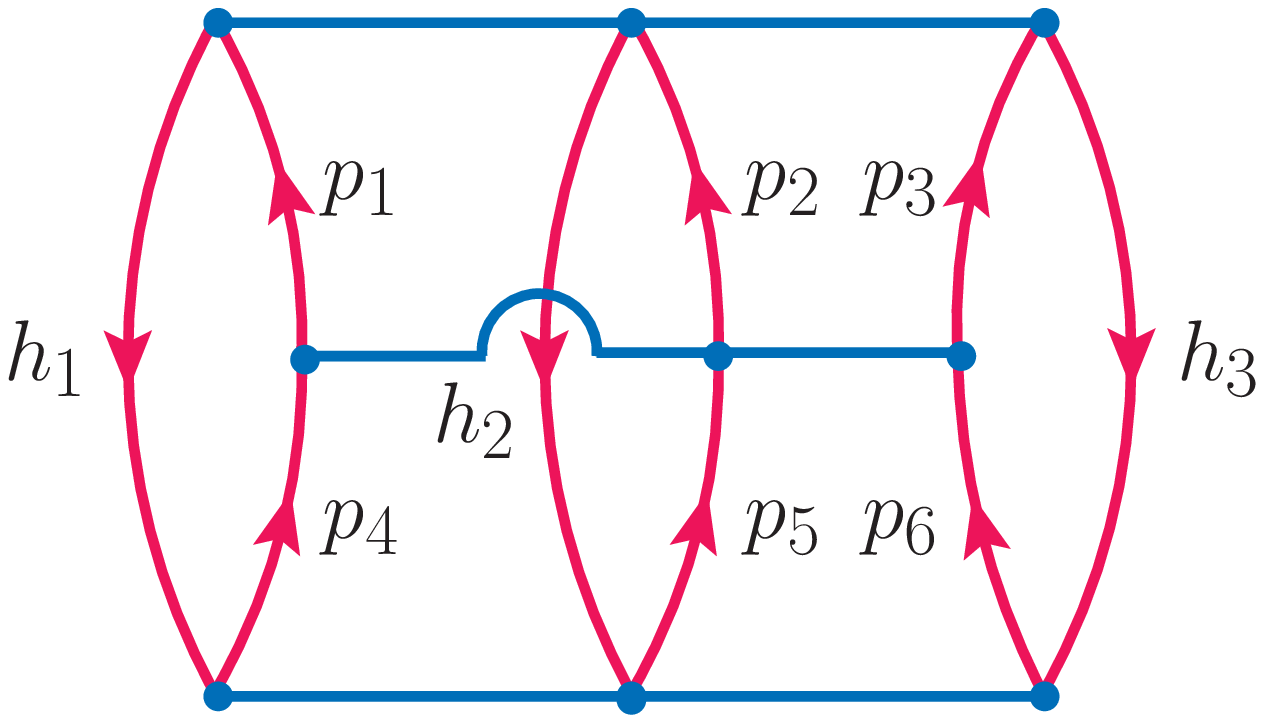}
}
}
}
\caption{The third-order ASG diagrams of energy corrections in the RSPT expansion.
The wavy line signifies the normal-ordered two-body Hamiltonian (\ref{DefineW})
which including the effects of 3$N$ interaction.
The solid line indicates the 3$N$ force.}
\label{fig:e3} 
\end{figure}

The RSPT energy up to third order according to the perturbation orders of $\hat{V}$
can be written as \cite{PhysRevC.94.014303}
\begin{equation}
E=E^{(0)}+E^{(1)}+E^{(2)}+E^{(3)}.
\end{equation}
In HF-RSPT, the summation of the $E^{(0)}$ and $E^{(1)}$ gives the HF energy $E_{\text{HF}}$,
\begin{equation}	
\label{HFenergymid}
\begin{split}
E^{(0)}+E^{(1)}=E_{\text{HF}}=\frac{1}{3} \sum_{i}\left(\varepsilon_{i}+2\langle i|\hat{H}^{(1)}|i \rangle \right)
+\frac{1}{6}\sum_{i,j}\langle ij|\hat{H}^{(2)}|ij \rangle
\end{split}
\end{equation}
with the HF single particle energies $\varepsilon_{p}$,
\begin{equation}	
\label{SingleStateEnergy}
\varepsilon_{p}=\langle p|\hat{H}^{(1)}|p \rangle+\sum_{i=1}^{A}\langle pi|\hat{H}^{(2)}|pi\rangle
+\frac{1}{2}\sum_{i,j=1}^{A}\langle pij|\hat{V}_{3N}|pij\rangle.
\end{equation}
The second-order energy correction reads
\begin{equation}	
\label{E2}
\begin{split}
\displaystyle E^{(2)}=
\dfrac{1}{4}\sum_{ij}\sum_{ab}
\dfrac{|\langle ij|\hat{W}|ab \rangle|^{2}}{\varepsilon_{i}+\varepsilon_{j}-\varepsilon_{a}-\varepsilon_{b}}
+\dfrac{1}{36}\sum_{ijk}\sum_{abc}
\dfrac{|\langle ijk|\hat{V}_{3N}|abc \rangle|^{2}}{\varepsilon_{i}+\varepsilon_{j}+\varepsilon_{k}-\varepsilon_{a}-\varepsilon_{b}-\varepsilon_{c} }
\end{split}.
\end{equation}
If we take $\hat{H}^{(2)}$ and $\hat{V}_{3N}$ as the RSPT vertices, there are 56 terms in the third-order energy corrections $E^{(3)}$ (see supplementary material).
While using $\hat{W}$ and $\hat{V}_{3N}$ as the RSPT vertices, there are only 17 terms.
Fig.~{\ref{fig:e3}} displays the anti-symmetrized Goldstone (ASG) diagrams corresponding to the 17 terms in third-order energy corrections.

\section{Calculations and discussions}
We now turn to the RSPT3 calculations with $NN$+3$N$ interaction in HF basis.
The HF is carried out within the HO basis,
and the HO basis is truncated by a cutoff according to
the number $N_{\rm shell} = \text{max}(2n+l +1)$,
where $N_{\rm shell}$ indicates how many major HO shells are included in the truncation.
We apply two different realistic $NN$ interactions plus a phenomenological contact 3$N$ force to the closed-shell nuclei, $^{4}$He, $^{16}$O and $^{40}$Ca.
We also focus on the state-of-the-art $\chi$NNLO$_{\rm sat}$ $NN$+3$N$ interaction \cite{PhysRevC.91.051301}
to investigate the structure of $^{4}$He, $^{14,22}$C, $^{16,22,24}$O and $^{40,48}$Ca.

\subsection{Calculations with the phenomenological three-body contact potential}

\begin{figure}
\centering
\includegraphics[scale=0.50]{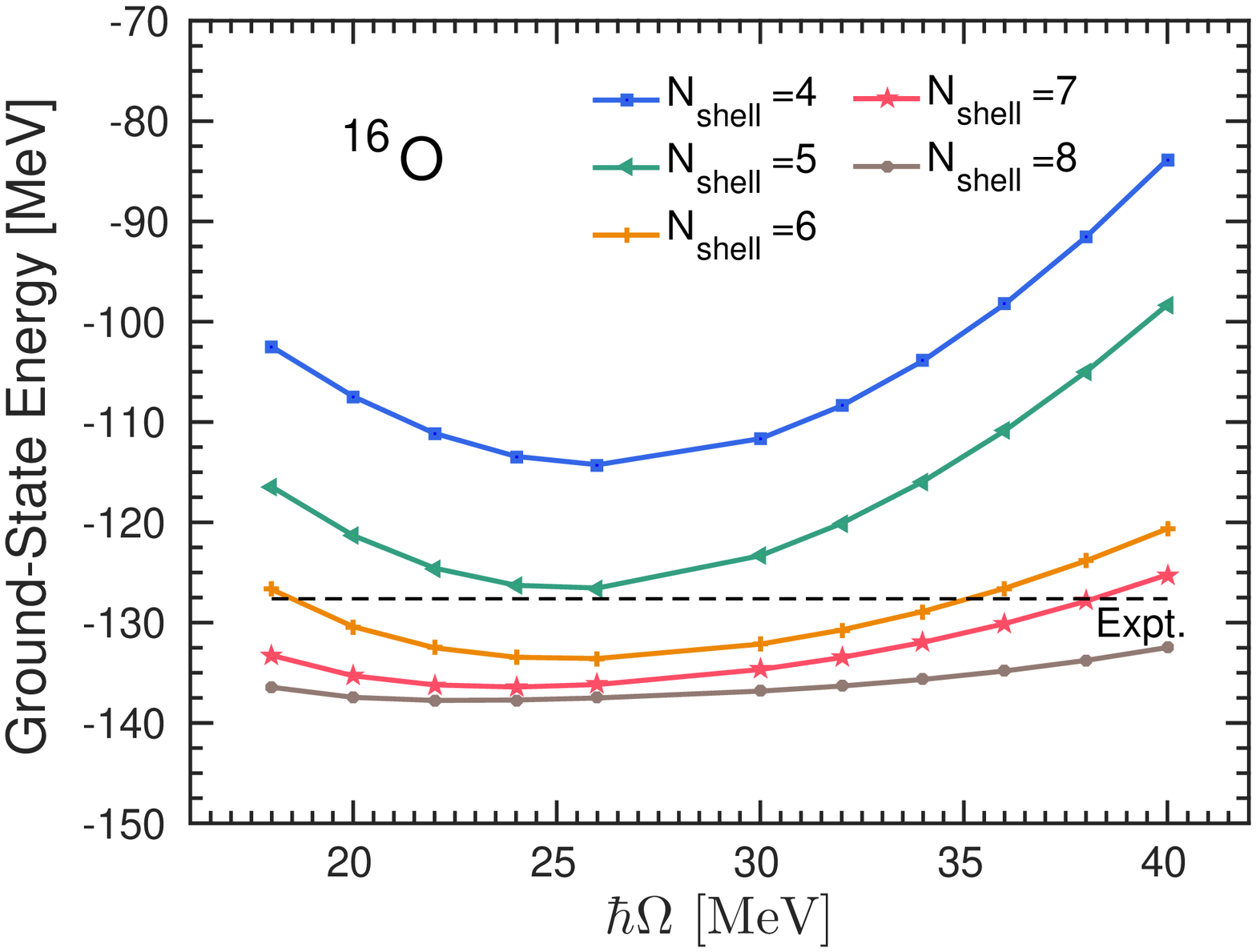}
\caption{\label{fig:RSPT_v2b_O16}
$^{16}\text{O}$ ground-state energy calculated by HF-RSPT through third order as a function of oscillator parameter $\hbar \Omega$.  The $\chi$N$^{3}$LO $NN$ interaction \cite{Machleidt20111,PhysRevC.68.041001} renormalized by $V_{\text{low-}k}$ at cutoff momentum $\Lambda=2.1$ $\text{fm}^{-1}$ is used. The dashed line represents the experimental ground-state energy.}
\end{figure}
\begin{figure}
\centering
\includegraphics[scale=0.50]{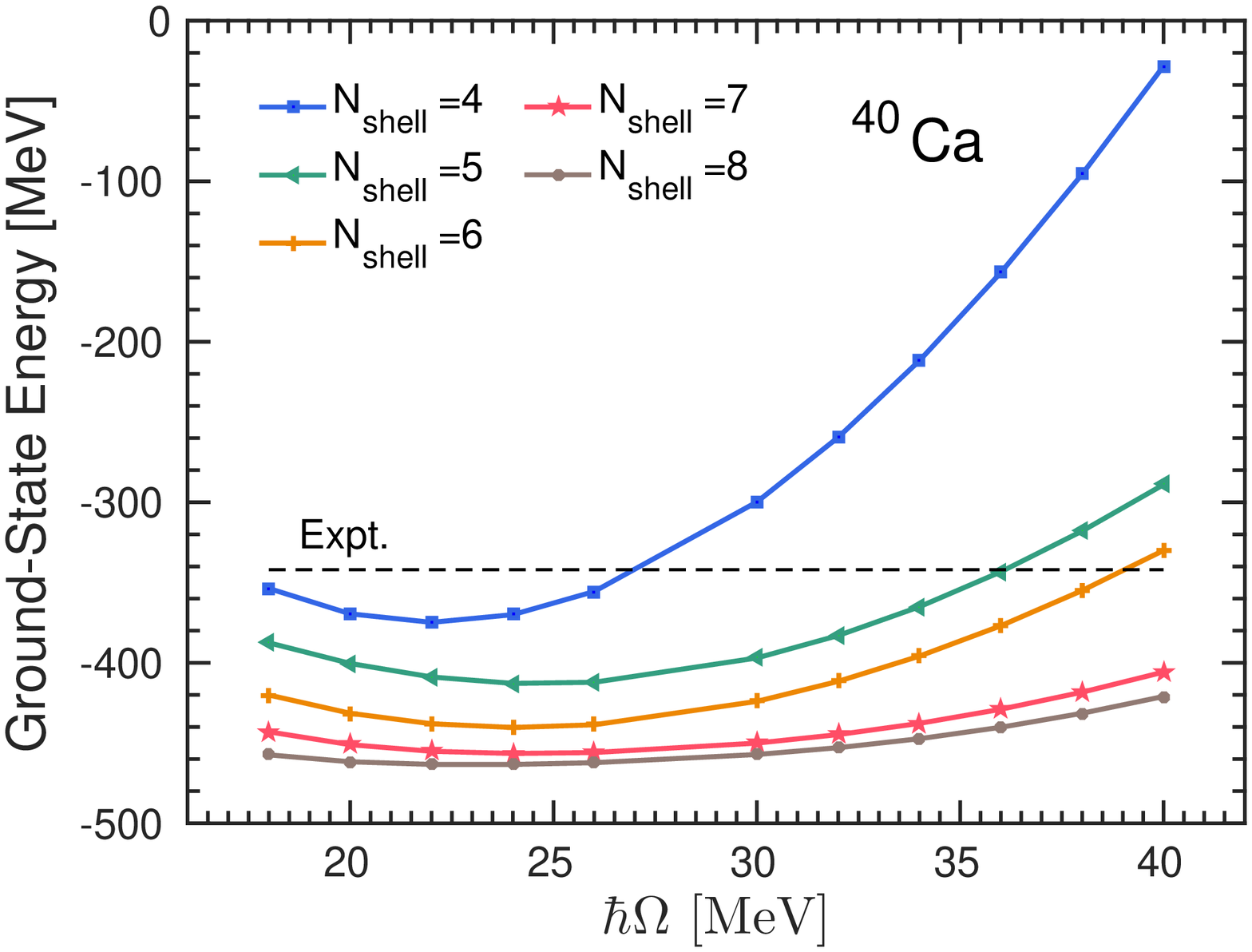}
\caption{\label{fig:RSPT_v2b_Ca40}
Similar to Fig. \ref{fig:RSPT_v2b_O16}, but for $^{40}$Ca.}
\end{figure}
A proof-of-principle calculation is given in the following by using softened $\chi$N$^{3}$LO \cite{PhysRevC.68.041001} and ``bare'' JISP16 $NN$ interactions \cite{PhysRevC.70.044005,Shirokov200596,Shirokov200733} plus phenomenological contact 3$N$ force \cite{PhysRevC.82.024319}. In the calculations, we take $N_{\rm shell}=7$ and $\hbar \Omega=30$ MeV for the basis space. This choice takes into account the balance between the convergence of the only $NN$ results and the great computational cost of 3$N$ force.
Figures \ref{fig:RSPT_v2b_O16} and \ref{fig:RSPT_v2b_Ca40} show
the two-body level HF-RSPT3 calculated ground-state energy of $^{16}$O and $^{40}$Ca as a function of $\hbar \Omega$ with different $N_{\rm shell}$.
The calculations are done with
the $\chi$N$^{3}$LO $NN$-only interactions softened by $V_{\text{low-}k}$ renormalization scheme \cite{PhysRevC.65.051301,Bogner20031}.
We took the most commonly used $V_{\text{low-}k}$ momentum cutoff $\Lambda=2.1$ fm${^{-1}}$ \cite{PhysRevC.65.051301,PhysRevC.71.014307}.
We see that good convergence of the calculated energy can be obtained even using a small $N_{\rm shell}=7$ and 8 truncations,
and well converged energy minima are found at $\hbar \Omega \approx 30$ MeV for all calculated nuclei.
A detailed analysis of the order-by-order convergence in HF-RSPT was presented in Ref. \cite{PhysRevC.94.014303}.
The basis convergence for the ``bare" JISP16 can refer to our previous work \cite{PhysRevC.94.014303}.
In that calculation we find that the  $N_{\text{shell}}=7$ and $\hbar \Omega=30$ MeV
is also a good choice.

\begin{figure}
\centering
\includegraphics[scale=0.65]{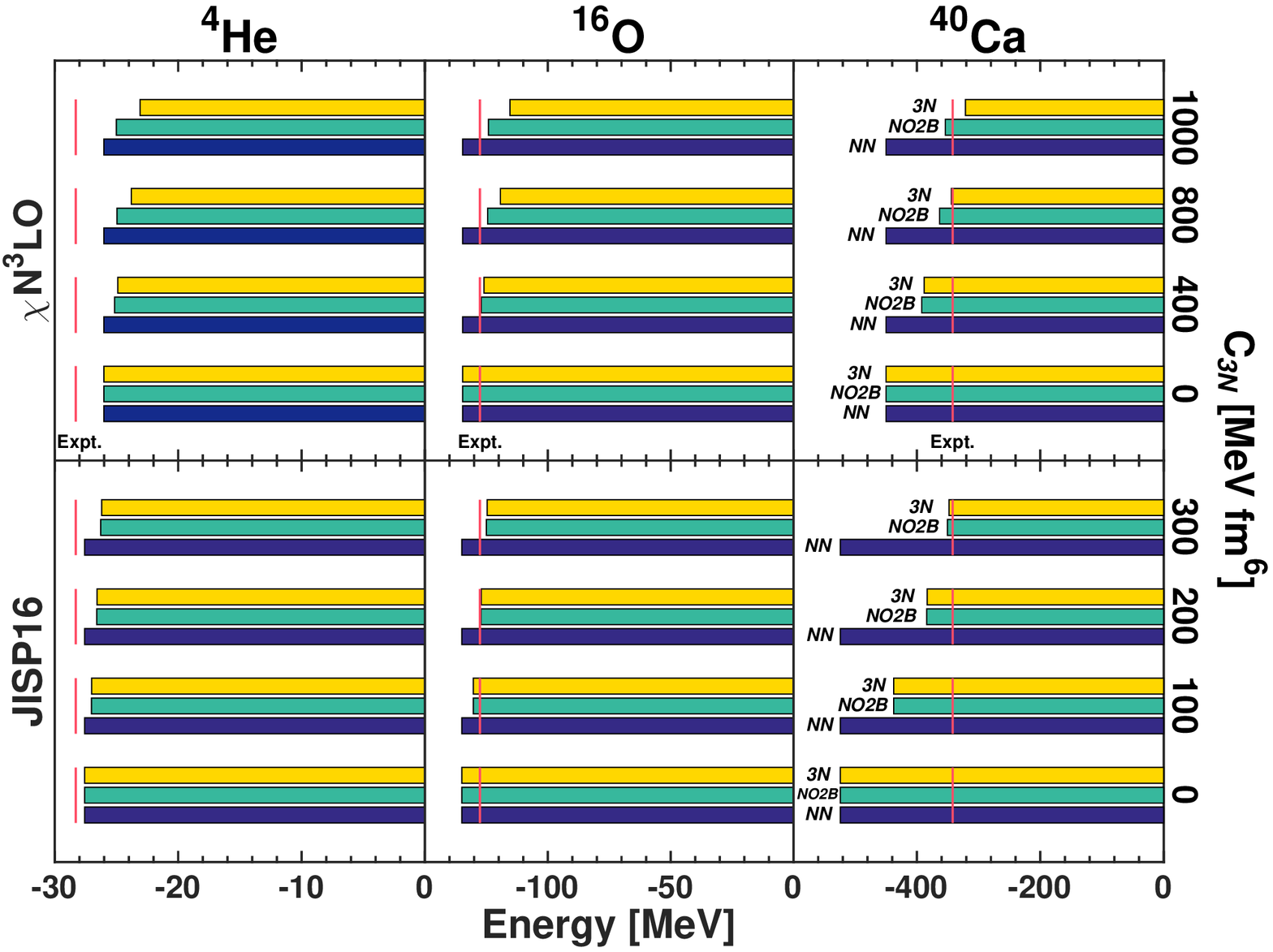}
\caption{\label{fig:C_3N} 
Ground-state energies of $^4$He, $^{16}$O and $^{40}$Ca computed at different normal-ordering level for $NN$ plus contact 3$N$ interaction, i.e., only $NN$ (indicated by $NN$), NO2B approximation for $NN$+3$N$ (indicated by NO2B) and full $NN$+3$N$ (indicated by 3$N$). The bar charts also show the calculations with different strength parameters $C_{3N}$ of 3$N$ contact potential. The upper panel gives the anatomy with the $V_{\text{low-}k}$-softened $\chi$N$^{3}$LO plus contact 3$N$ potential. While the lower panel shows the results with two-body JISP16 plus contact 3$N$ interaction.}
\end{figure}
The impact of 3$N$ contact interaction with different strength parameters $C_{3N}$
is illustrated in Fig. \ref{fig:C_3N}.
In the upper panel the $\chi$N$^3$LO $NN$ interaction are used, while in the lower panel the $NN$ interaction is the JISP16 .
We see that the $NN$-only interactions give an overbinding compared to experiment
for $^{16}$O and $^{40}$Ca but $^{4}$He.
The purely repulsive contact 3$N$ force can improve energies in $^{16}$O and $^{40}$Ca.
The contact 3$N$ force over a large $C_{3N}$ range has little effect on $^{4}$He,
while gives an increasing contribution from $^{16}$O to $^{40}$Ca.
This implies that the effects of 3$N$ force will become more pronounced on the heavier nuclei.
The fitted $C_{3N}\approx 300$  and 800 $\text{MeV} \cdot \text{fm}^{6}$ in $\chi$N$^3$LO evaluations
can reproduce the binding energy of $^{16}$O and $^{40}$Ca, respectively.
While $C_{3N}\approx 200$  and 320 $\text{MeV} \cdot \text{fm}^{6}$ in JISP16  calculations.
For JISP16 the $C_{3N}$ is smaller than $\chi$N$^3$LO.
This is because that the JISP16 potential can minimize the need of 3$N$ force by phase-equivalent transformation of off-shell freedom. Polyzou and Glockle \cite{latePolyzou} have shown that changing the off-shell properties of the two-body potential is equivalent to adding many-body interactions.
Comparing the NO2B approximation with the full 3$N$ calculation, they are very similar near the fit point for all Hamiltonian cases.

\subsection{Calculations with chiral NNLO$_{\rm sat}$ potential}
\begin{table}
\centering
\caption{
\label{be}
Ground-state energies (in MeV) of $^4$He, $^{14,22}$C, $^{16,22,24}$O and $^{40,48}$Ca
computed within HF-RSPT, compared to the CC, IM-SRG and experimental data.
The $\chi$NNLO$_{\rm sat}$ interactions with $N_{\text{shell}}=13$ and $\hbar\Omega=22$ MeV are used in calculations.
}
\begin{tabular}{ccccc p{cm}}
\hline
\hline
Nucleus & CC~\cite{PhysRevC.91.051301} & IM-SRG & HF-RSPT3  &  Expt.~\cite{ame2016} \\
\hline
$^4$He    &$-$28.43 &$-$29.09 &$-$28.15 &$-$28.30\\
$^{14}$C  &$-$103.6 &$-$104.16&$-$105.06&$-$105.29\\
$^{22}$C  &$-$      &$-$114.79&$-$113.95&$-$119.18\\
$^{16}$O  &$-$124.4 &$-$124.13&$-$125.16&$-$127.62\\
$^{22}$O  &$-$160.8 &$-$160.02&$-$156.77&$-$162.03\\
$^{24}$O  &$-$168.1 &$-$166.26&$-$163.12&$-$168.97\\
$^{40}$Ca &$-$326.0      &$-$311.47&$-$320.66&$-$342.05\\
$^{48}$Ca &$-$      &$-$376.75&$-$370.02&$-$416.00\\
\hline
\hline
\end{tabular}
\end{table}
\begin{table}
\centering
\caption{
\label{radii}
Similar to Table~\ref{be}, but for charge radii (in fm). HF-RSPT1 designates the leading-order HF-RSPT, i.e., HF calculation.
}
\begin{tabular}{ccccc p{cm}}
\hline
\hline
\multicolumn{1}{c}{} & \multicolumn{1}{c}{CC~\cite{PhysRevC.91.051301}} &
\multicolumn{1}{c}{IM-SRG} & \multicolumn{1}{c}{HF-RSPT1} & \multicolumn{1}{c}{Expt.~\cite{Angeli201369}}\\
\hline
$^4$He    &1.70&1.69&1.75&1.6755(28)\\
$^{14}$C  &2.48&2.43&2.57&2.5025(87)\\
$^{22}$C  &$-$ &2.53&2.63&$-$\\
$^{16}$O  &2.71&2.67&2.78&2.6991(52)\\
$^{22}$O  &2.72 &2.66&2.75&$-$\\
$^{24}$O  &2.76&2.70&2.78&$-$\\
$^{40}$Ca &3.48&3.40&3.49&3.4776(19)\\
$^{48}$Ca &$-$ &3.38&3.46&3.4771(20)\\
\hline
\hline
\end{tabular}
\end{table}
After validating the NO2B approximation in the HF-RSPT3,
we are now applying it to the {\it ab initio} calculations with $\chi$NNLO$_{\rm sat}$ $NN$+3$N$ interaction.
Table \ref{be} gives the ground-state energies of $^{4}$He, $^{14,22}$C, $^{16,22,24}$O and $^{40,48}$Ca calculated by HF-RSPT3 .
The underlying HO basis parameter $\hbar \Omega=22$ MeV is taken,
and the basis is truncated with $N_{\rm shell}=13$.
We see that the HF-RSPT3 energies with $NN$+3$N$ interaction are in overall agreement with data.
We also compare the results with IM-SRG \cite{PhysRevC.93.051301} and CC \cite{PhysRevC.91.051301} evaluations.
The IM-SRG are performed using new Magnus formulation to decouple the Hamiltonians \cite{PhysRevC.92.034331}.
The results of HF-RSPT3 are in 2$\%$-level agreement with the nonperturbative IM-SRG.
These binding energies are also in good agreement with the coupled-cluster $\Lambda$-CCSD(T) calculations \cite{PhysRevC.91.051301}.
For the weakly-bound $^{22}$C, we underestimate the binding energy because of lacking continuum coupling \cite{Hu:2018rnc}.

Table \ref{radii} shows the computed charge radii compared to the data.
In Ref. \cite{PhysRevC.94.014303}, we have concluded that the second-order radius correction is very small in HF-RSPT, and the leading-order perturbation (i.e., HF calculation) can catch the main contribution.
So we calculate the HF-RSPT radius only in leading-order level. The IM-SRG gives radius by deriving an effective point-proton radius operator \cite{HERGERT2016165}.
Compared to the result of IM-SRG and CC, the HF-RSPT1 gives similar charge radius.
These {\it ab initio} calculations with $\chi$NNLO$_{\text{sat}}$ $NN$+3$N$ interaction give good descriptions of nuclear bulk properties (including their masses and radii) from light- to medium-mass regions.

\section{Conclusions}
In conclusion, we have developed a third-order Rayleigh-Schr\"{o}dinger perturbation theory (RSPT3) with three-body interaction for the first time.
Starting from two-body realistic nuclear forces (i.e., chiral $\mathrm{N^{3}LO}$ and JISP16) and phenomenological three-body contact potential, we have calculated the structure of $\mathrm{^{4}He}$, $\mathrm{^{16}O}$, and $\mathrm{^{40}Ca}$ by performing RSPT3 within Hartree-Fock (HF) bases (HF-RSPT3). Compared with the results that only use two-body force ($NN$), the inclusion of three-nucleon interaction (3$N$) can improve the calculation, giving good agreement with the experimental data.
Our results show that the main contribution of the normal-ordered 3$N$ interaction stems from the zero-, one- and two-body terms in RSPT, as IT-NCSM and CC claim. This implies that the NO2B approximation works very well in the RSPT.
To check the convergence of the RSPT calculation, we have made comparisons with benchmarks given by the best available IM-SRG and CC calculations with the same NO2B $\chi$NNLO$_{\text{sat}}$ potential.
The HF-RSPT3 gives 2$\%$-level agreement with the IM-SRG and CC calculations at a small fraction of the computational cost.
The three $\chi$NNLO$_{\text{sat}}$ calculations in $^{4}$He, $^{14,22}$C, $^{16,22,24}$O and $^{40,48}$Ca can reproduce binding energies and radii simultaneously.
Because of its low computational cost, the HF-RSPT3 provides a valuable and efficient tool to exploit full 3$N$ interaction exactly in medium-mass and heavy nuclei where the advanced non-perturbative methods are computationally too demanding.

\section*{Acknowledgements}

We are grateful to G.R. Jansen for producing the normal-ordered matrix elements of the interaction NNLO$_{\rm sat}$ in the Hartree-Fock basis.
This work has been supported by
the National Key R${\&}$D Program of China under Grant No. 2018YFA0404401;
the National Natural Science Foundation of China under Grants No. 11835001, No. 11320101004 and No. 11575007;
the China Postdoctoral Science Foundation under Grant No. 2018M630018;
and the CUSTIPEN (China-U.S. Theory Institute for Physics with Exotic Nuclei) funded by the U.S.  Department of Energy,
Office of Science under Grant No. DE-SC0009971.
We acknowledge the High-performance Computing Platform of Peking University for providing computational resources.
The IM-SRG code used is Ragnar$\_$IMSRG \cite{ragnar}.


\section*{References}

  \bibliographystyle{elsarticle-num_noURL}
  \bibliography{references}

\begin{thebibliography}{10}
\expandafter\ifx\csname url\endcsname\relax
  \def\url#1{\texttt{#1}}\fi
\expandafter\ifx\csname urlprefix\endcsname\relax\def\urlprefix{URL }\fi
\expandafter\ifx\csname href\endcsname\relax
  \def\href#1#2{#2} \def\path#1{#1}\fi

\bibitem{Machleidt20111}
R.~Machleidt, D.~R. Entem,
  \href{http://www.sciencedirect.com/science/article/pii/S0370157311000457}{Physics
  Reports} 503~(1) (2011) 1 -- 75.

\bibitem{PhysRevC.86.054317}
F.~Sammarruca, B.~Chen, L.~Coraggio, N.~Itaco, R.~Machleidt,
  \href{https://link.aps.org/doi/10.1103/PhysRevC.86.054317}{Phys. Rev. C} 86
  (2012) 054317.

\bibitem{PhysRevC.89.044321}
L.~Coraggio, J.~W. Holt, N.~Itaco, R.~Machleidt, L.~E. Marcucci, F.~Sammarruca,
  \href{https://link.aps.org/doi/10.1103/PhysRevC.89.044321}{Phys. Rev. C} 89
  (2014) 044321.

\bibitem{PhysRevC.91.051301}
A.~Ekstr\"om, G.~R. Jansen, K.~A. Wendt, G.~Hagen, T.~Papenbrock, B.~D.
  Carlsson, C.~Forss\'en, M.~Hjorth-Jensen, P.~Navr\'atil, W.~Nazarewicz,
  \href{http://link.aps.org/doi/10.1103/PhysRevC.91.051301}{Phys. Rev. C} 91
  (2015) 051301.

\bibitem{PhysRevC.70.044005}
A.~M. Shirokov, A.~I. Mazur, S.~A. Zaytsev, J.~P. Vary, T.~A. Weber,
  \href{http://link.aps.org/doi/10.1103/PhysRevC.70.044005}{Phys. Rev. C} 70
  (2004) 044005.

\bibitem{PhysRevLett.103.082501}
E.~D. Jurgenson, P.~Navr\'{a}til, R.~J. Furnstahl,
  \href{http://link.aps.org/doi/10.1103/PhysRevLett.103.082501}{Phys. Rev.
  Lett.} 103 (2009) 082501.

\bibitem{PhysRevLett.106.202502}
P.~Maris, J.~P. Vary, P.~Navr\'atil, W.~E. Ormand, H.~Nam, D.~J. Dean,
  \href{https://link.aps.org/doi/10.1103/PhysRevLett.106.202502}{Phys. Rev.
  Lett.} 106 (2011) 202502.

\bibitem{PhysRevLett.110.022502}
J.~D. Holt, J.~Men\'endez, A.~Schwenk,
  \href{https://link.aps.org/doi/10.1103/PhysRevLett.110.022502}{Phys. Rev.
  Lett.} 110 (2013) 022502.

\bibitem{PhysRevLett.110.242501}
H.~Hergert, S.~Binder, A.~Calci, J.~Langhammer, R.~Roth,
  \href{https://link.aps.org/doi/10.1103/PhysRevLett.110.242501}{Phys. Rev.
  Lett.} 110 (2013) 242501.

\bibitem{PhysRevLett.113.262504}
A.~Ekstr\"om, G.~R. Jansen, K.~A. Wendt, G.~Hagen, T.~Papenbrock, S.~Bacca,
  B.~Carlsson, D.~Gazit,
  \href{https://link.aps.org/doi/10.1103/PhysRevLett.113.262504}{Phys. Rev.
  Lett.} 113 (2014) 262504.

\bibitem{RevModPhys.87.1067}
J.~Carlson, S.~Gandolfi, F.~Pederiva, S.~C. Pieper, R.~Schiavilla, K.~E.
  Schmidt, R.~B. Wiringa,
  \href{https://link.aps.org/doi/10.1103/RevModPhys.87.1067}{Rev. Mod. Phys.}
  87 (2015) 1067--1118.

\bibitem{RevModPhys.85.197}
H.-W. Hammer, A.~Nogga, A.~Schwenk,
  \href{https://link.aps.org/doi/10.1103/RevModPhys.85.197}{Rev. Mod. Phys.} 85
  (2013) 197--217.

\bibitem{PhysRevC.76.034302}
G.~Hagen, T.~Papenbrock, D.~J. Dean, A.~Schwenk, A.~Nogga, M.~W\l{}och,
  P.~Piecuch, \href{https://link.aps.org/doi/10.1103/PhysRevC.76.034302}{Phys.
  Rev. C} 76 (2007) 034302.

\bibitem{PhysRevLett.109.052501}
R.~Roth, S.~Binder, K.~Vobig, A.~Calci, J.~Langhammer, P.~Navr\'atil,
  \href{https://link.aps.org/doi/10.1103/PhysRevLett.109.052501}{Phys. Rev.
  Lett.} 109 (2012) 052501.

\bibitem{Rayleigh1894}
J.~W.~S. Baron~Rayleigh, Theory of sound, Dover Publications, second edition,
  Vol.1, 1894.

\bibitem{Schrodinger1926}
E.~Schr{\"o}dinger, Ann. Phys. (Leipzig) 385 (1926) 437--490.

\bibitem{bartlett2009}
I.~Shavitt, R.~J. Bartlett, Many-Body Methods in Chemistry and Physics: MBPT
  and Coupled-Cluster Theory, Cambridge University Press, 2009.

\bibitem{PhysRevC.95.034321}
B.~S. Hu, F.~R. Xu, Q.~Wu, Y.~Z. Ma, Z.~H. Sun,
  \href{https://link.aps.org/doi/10.1103/PhysRevC.95.034321}{Phys. Rev. C} 95
  (2017) 034321.

\bibitem{PhysRevC.68.034320}
L.~Coraggio, N.~Itaco, A.~Covello, A.~Gargano, T.~T.~S. Kuo,
  \href{http://link.aps.org/doi/10.1103/PhysRevC.68.034320}{Phys. Rev. C} 68
  (2003) 034320.

\bibitem{PhysRevC.69.034332}
M.~A. Hasan, J.~P. Vary, P.~Navr\'atil,
  \href{http://link.aps.org/doi/10.1103/PhysRevC.69.034332}{Phys. Rev. C} 69
  (2004) 034332.

\bibitem{PhysRevC.73.044312}
R.~Roth, P.~Papakonstantinou, N.~Paar, H.~Hergert, T.~Neff, H.~Feldmeier,
  \href{http://link.aps.org/doi/10.1103/PhysRevC.73.044312}{Phys. Rev. C} 73
  (2006) 044312.

\bibitem{1674-1137-41-10-104101}
B.~S. Hu, Q.~Wu, F.~R. Xu,
  \href{http://stacks.iop.org/1674-1137/41/i=10/a=104101}{Chinese Physics C}
  41~(10) (2017) 104101.

\bibitem{PhysRevLett.99.092501}
R.~Roth, P.~Navr\'atil,
  \href{https://link.aps.org/doi/10.1103/PhysRevLett.99.092501}{Phys. Rev.
  Lett.} 99 (2007) 092501.

\bibitem{Tichai:2018mll}
A.~Tichai, P.~Arthuis, T.~Duguet, H.~Hergert, V.~Som\`a, R.~Roth,
  arXiv:1806.10931 [nucl-th] (2018).

\bibitem{PhysRevC.94.014303}
B.~S. Hu, F.~R. Xu, Z.~H. Sun, J.~P. Vary, T.~Li,
  \href{https://link.aps.org/doi/10.1103/PhysRevC.94.014303}{Phys. Rev. C} 94
  (2016) 014303.

\bibitem{TICHAI2016283}
A.~Tichai, J.~Langhammer, S.~Binder, R.~Roth,
  \href{http://www.sciencedirect.com/science/article/pii/S0370269316002008}{Physics
  Letters B} 756 (2016) 283 -- 288.

\bibitem{PhysRevC.68.041001}
D.~R. Entem, R.~Machleidt,
  \href{http://link.aps.org/doi/10.1103/PhysRevC.68.041001}{Phys. Rev. C} 68
  (2003) 041001.

\bibitem{Shirokov200596}
A.~Shirokov, J.~Vary, A.~Mazur, S.~Zaytsev, T.~Weber,
  \href{http://www.sciencedirect.com/science/article/pii/S0370269305008518}{Physics
  Letters B} 621~(1-2) (2005) 96 -- 101.

\bibitem{Shirokov200733}
A.~M. Shirokov, J.~P. Vary, A.~I. Mazur, T.~A. Weber,
  \href{http://www.sciencedirect.com/science/article/pii/S0370269306014158}{Physics
  Letters B} 644~(1) (2007) 33 -- 37.

\bibitem{PhysRevC.82.024319}
A.~G\"unther, R.~Roth, H.~Hergert, S.~Reinhardt,
  \href{https://link.aps.org/doi/10.1103/PhysRevC.82.024319}{Phys. Rev. C} 82
  (2010) 024319.

\bibitem{PhysRevC.91.044001}
K.~Hebeler, H.~Krebs, E.~Epelbaum, J.~Golak, R.~Skibi\ifmmode~\acute{n}\else
  \'{n}\fi{}ski,
  \href{http://link.aps.org/doi/10.1103/PhysRevC.91.044001}{Phys. Rev. C} 91
  (2015) 044001.

\bibitem{PhysRevC.65.051301}
S.~K. Bogner, T.~T.~S. Kuo, L.~Coraggio, A.~Covello, N.~Itaco,
  \href{http://link.aps.org/doi/10.1103/PhysRevC.65.051301}{Phys. Rev. C} 65
  (2002) 051301.

\bibitem{Bogner20031}
S.~K. Bogner, T.~T.~S. Kuo, A.~Schwenk,
  \href{http://www.sciencedirect.com/science/article/pii/S0370157303002953}{Physics
  Reports} 386~(1) (2003) 1 -- 27.

\bibitem{PhysRevC.71.014307}
L.~Coraggio, A.~Covello, A.~Gargano, N.~Itaco, T.~T.~S. Kuo, R.~Machleidt,
  \href{http://link.aps.org/doi/10.1103/PhysRevC.71.014307}{Phys. Rev. C} 71
  (2005) 014307.

\bibitem{latePolyzou}
W.~Polyzou, W.~Gl\"{o}ckle,
  \href{http://dx.doi.org/10.1007/BF01091701}{Few-Body Systems} 9~(2-3) (1990)
  97--121.

\bibitem{ame2016}
M.~Wang, G.~Audi, F.~G. Kondev, W.~J. Huang, S.~Naimi, X.~Xu,
  \href{http://cpc-hepnp.ihep.ac.cn:8080/Jwk_cpc/EN/abstract/article_8345.shtml}{Chinese
  physics C} 41~(3) (2017) 30003.

\bibitem{Angeli201369}
I.~Angeli, K.~Marinova,
  \href{http://www.sciencedirect.com/science/article/pii/S0092640X12000265}{Atomic
  Data and Nuclear Data Tables} 99~(1) (2013) 69 -- 95.

\bibitem{PhysRevC.93.051301}
S.~R. Stroberg, H.~Hergert, J.~D. Holt, S.~K. Bogner, A.~Schwenk,
  \href{https://link.aps.org/doi/10.1103/PhysRevC.93.051301}{Phys. Rev. C} 93
  (2016) 051301.

\bibitem{PhysRevC.92.034331}
T.~D. Morris, N.~M. Parzuchowski, S.~K. Bogner,
  \href{https://link.aps.org/doi/10.1103/PhysRevC.92.034331}{Phys. Rev. C} 92
  (2015) 034331.

\bibitem{Hu:2018rnc}
B.~S. Hu, Q.~Wu, Z.~H. Sun, F.~R. Xu, arXiv:1809.08405 [nucl-th].

\bibitem{HERGERT2016165}
H.~Hergert, S.~Bogner, T.~Morris, A.~Schwenk, K.~Tsukiyama,
  \href{http://www.sciencedirect.com/science/article/pii/S0370157315005414}{Physics
  Reports} 621 (2016) 165 -- 222, memorial Volume in Honor of Gerald E. Brown.

\bibitem{ragnar}
S.~R. Stroberg, https://github.com/ragnarstroberg/ragnar\_imsrg.

\end{thebibliography}





\end{document}